# Optical Properties of Aluminium-Doped Zinc Oxide Thin Films Synthesized via AACVD Using Nitrogen as a Carrier Gas

Kingsley Imoni-Ogbe*[1], Onyekachukwu Mike Osiele[2], Vincent Akpoveta[3], Queen Umudi[4], Bright Ugbolu[5], Oscar Enajite[6]

*[1]Department of Physics, University of Delta, Agbor, Delta State, Nigeria.
[2]Department of Physics, Delta State University, Abraka, Delta State, Nigeria.
[3]Department of Chemical Sciences, Dennis Osadebay University, Asaba, Delta State, Nigeria.
[4]Deprtment of Chemistry, University of Delta, Agbor, Delta State, Nigeria.
[5]Department of Chemistry and Biochemistry, Florida State University, Tallahasse, Fl, 323.6, USA
[6]Department of Physical Sciences, Dennis Osadebay University, Asaba, Delta State, Nigeria.

*[1]Corresponding Author Email: kingsley.imoni-ogbe@unidel.edu.ng, kudos12j@gmail.com

## Abstract

This research combines AACVD technology with $N_2$ carrier gas to produce AZO thin films while studying the structural and optical and morphological effects of Al doping at concentrations ranging from 0% to 20%. The AACVD process took 400°C to deposit films onto soda-lime glass surfaces before an annealing step at 450°C under nitrogen environment.

When utilizing nitrogen gas as carrier medium in X-ray Diffraction tests showed superior crystalline structure through distinct powerful signals appearing in the (220), (311), and (222) peaks. The films demonstrated smaller crystallite sizes and greater dislocation densities when the Al doping concentration increased because this produced more lattice strain as well as structural defects. The thickness of films obtained through profilometry increased slightly from undoped ZnO at 102 nm to 115 nm for 20% Al-doped ZnO samples.

The UV absorbance of nitrogen-annealed films became stronger with peak intensities being more pronounced at lower doping stages (5%). The bandgap energy of the produced films expanded steadily when Al concentrations increased from undoped ZnO at 3.21 eV to 3.33 eV for ZnO with 20% Al doping. This change matched the Burstein-Moss effect. Optoelectronic devices benefit from the enhanced optical activity over broader wavelength ranges through the Al doping rates exceeding 15% and up to 20% levels in these films.

The research reveals nitrogen carrier gas plays a substantial role in shaping AZO thin film crystal properties while increasing dislocation density numbers and resulting in slightly heavier film thickness than oxygen gas.

**Keywords:** Aluminium-doped Zinc Oxide (AZO), Optical properties, Aerosol-Assisted Chemical Vapor Deposition (AACVD), Nitrogen carrier gas and Bandgap

## 1.0 Introduction

AZO films composed of aluminium-doped zinc oxide have become extensively studied in the optoelectronic field because of their ability to be simultaneously transparent and conductive. The optical characteristics of AZO as a transparent conductive oxide (TCO) allow it to be an alternative to indium tin oxide (ITO) because AZO enables excellent visible-range optical transmittance and high electrical conductivity (Minami, 2005). The zinc oxide (ZnO) shows improved electrical properties after aluminium (Al) doping because the free electron concentration increases significantly for applications that require such properties like solar cells and light-emitting diodes (LEDs) and transparent display electrodes (Ozgur et al., 2005). The advantages of AZO films relate to their stability combined with their environmentally friendly features and economical characteristics which make them suitable for many optoelectronic devices (Kołodziejczak-Radzimska & Jesionowski, 2014). Advance in next-generation optoelectronic applications depends on the ability to control AZO film features through doping methods and AACVD deposition techniques (Wang et al., 2012).

Through doping processes zinc oxide (ZnO) films gain highly desirable changes in their structural characteristics together with optical and electrical characteristics which enable their application in many optoelectronic systems. Aluminium (Al) doping in ZnO enhances the electrical conductivity of the material substantially because it creates free charge carriers (Ozgur et al., 2005). Zinc oxide crystal property improvement occurs because zinc atoms in the material are replaced with aluminium donors that boost electron levels in the conduction band (Minami, 2005). When applying doping to ZnO films the resultant bandgap widens because carriers increase the Fermi level that shifts to the conduction band according to Burstein-Moss effect (Janotti & Van de Walle, 2009). The bandgap adjustable nature serves applications best among solar cells and photodetectors since optimal light absorption along with carrier transport stands as critical requirements (Kołodziejczak-Radzimska & Jesionowski, 2014). Higher doping levels cause ZnO films to show decreased crystallite sizes and greater lattice strain which results in performance reduction of the material (Wang et al., 2012). Doping serves as a powerful method to adjust ZnO film characteristics which specifically match performance needs of modern optoelectronic devices.

Aerosol-assisted chemical vapour deposition (AACVD) establishes itself as a flexible method to produce top-quality thin films specifically aluminum-doped zinc oxide (AZO) films. The synthesis process starts with breaking a precursor solution down into droplet formations before the carrier gas shifts these droplets toward a heated surface where thermal decomposition results in the formation of a thin film (Parkin & Palgrave, 2005). AACVD demonstrates exceptional capabilities for producing uniform and homogeneous films across extensive areas therefore granting its suitability in industrial environments (Shinde et al., 2011). The AACVD processing method enables scientists to monitor three main deposition control variables including precursor concentration and carrier gas flow rate and substrate temperature that enable the precise adjustment of film characteristics such as film thickness and crystal arrangement and surface shape (Wang et al., 2012). The procedure exhibits versatility because it accepts many precursor components and transportation agents while nitrogen gas enables filmmakers to alter both structural character and optical attributes of the resulting coatings (Janotti & Van de Walle, 2009). The cost-efficient and environmentally sound nature of AACVD stands higher than other deposition methods because it works under lower temperature settings and avoids the need for expensive vacuum equipment (Kołodziejczak-Radzimska & Jesionowski, 2014). These advantages make AACVD a preferred choice for the synthesis of functional thin films for optoelectronic and photovoltaic applications.

The selection of carrier gas in Aerosol-Assisted Chemical Vapor Deposition (AACVD) determines both the film growth characteristics and final thin film properties. The inert properties of $N_2$ make it suitable for serving as a carrier gas in AACVD because they guard against unguided chemical interactions during the deposition run. The function of nitrogen serves two purposes: it conducts precursor vapor to the heated substrate surface for deposition to take place. The usage of non-reactive properties prevents modification of chemical composition in AZO films because they protect the incorporation of aluminium as doping material. Nitrogen stands as the preferred transport medium since it enables accurate structural and property manipulation during film production. The use of nitrogen as carrier gas yields superior results over oxygen by avoiding the introduction of oxygen species into deposited materials because this maintains their stability and uniformity of doping distribution. Equipped with nitrogen gas cells form thinner layers containing smaller crystallites and more dislocations (Shinde et al., 2011). The scarcity of reactive oxygen species within the film medium prevents the grain growth mechanism which impacts the film structure. The implementation of nitrogen leads to modifications in film thickness and morphology in addition to affecting their UV region absorption properties due to increased defect numbers (Ozgur et al., 2005). The flow rate and pressure settings of the carrier gas system enable the control of film properties including optical features and electrical values so nitrogen remains an essential choice for producing AZO thin films particularly in optoelectronics and sensor applications.

## 2. METHODOLOGY

### 2.1 Research Design

The research experiments with AACVD to create AZO thin films as they study how $N_2$ carrier gas affects film physical performance. The experiment investigates five types of Aluminium (Al) dopants ranging from 0% to 5% to 10% to 15% and up to 20% for understanding film structural and optical properties. Five test samples received nitrogen as carrier gas for the processes which were named S1N through S5N according to their aluminum composition.

**Table 2.1: List of samples**

| Sample | Dopant % Conc. | Carrier Gas | Annealing gas | Deposition T°C | Annealing T°C |
|---|---|---|---|---|---|
| S1N | 0 | $N_2$ | $N_2$ | 400 | 450 |
| S2N | 5 | $N_2$ | $N_2$ | 400 | 450 |
| S3N | 10 | $N_2$ | $N_2$ | 400 | 450 |
| S4N | 15 | $N_2$ | $N_2$ | 400 | 450 |
| S5N | 20 | $N_2$ | $N_2$ | 400 | 450 |

The samples named S1N, S2N, S3N, S4N, and S5N represent ZnO thin films containing Aluminium (Al) which received both carrier gas and annealing gas treatment with $N_2$. The experimental team deposited annealed the samples at controlled temperature conditions to evaluate the effects nitrogen had on morphological structure as well as crystal clarity and total operating performance.

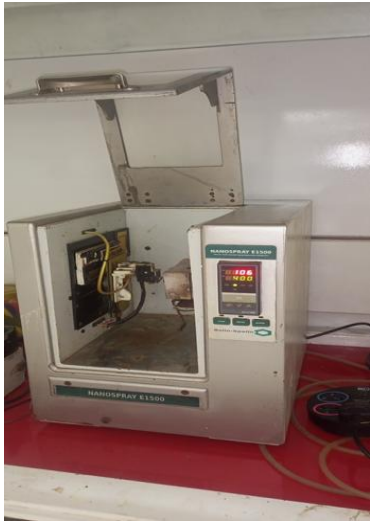

Plate 1: Aerosol Assisted Chemical Vapour Deposition (AACVD) Machine

The applied AACVD setup maintains conditions that allowAZO thin film deposition with exactness and continuous precision. A temperature of 400°C operates during the process while carrier gases ($O_2$ or $N_2$) flow at 3L/min and pressurize at 2.5 bar. The delivery of precursor solution containing 0.15M Zinc Acetylacetonate with 0.15M Aluminium Chloride in Methanol takes place through a 0.1ml/min flow rate. An atomized precursor occurs when the system applies 6kV voltage and positions the nozzle at 12mm from the substrate. Precursor solution of 0.3 milliliters is sprayed onto the samples before process annealing at 450 degrees Celsius occurs in nitrogen atmosphere for sixty minutes (Janotti and Van de Walle, 2009). The system enables researchers to manipulate deposition conditions to evaluate how carrier gas affects the properties of AZO films.

## 2.2 Materials and Equipment

Zinc Acetylacetonate along with Aluminium Chloride worked as precursors for Aluminium-doped Zinc Oxide (AZO) Thin Films synthesis while Methanol served as the solvent. The experimental deposition used carrier gas Nitrogen ($N_2$) in order to study its effects on the final film characteristics.

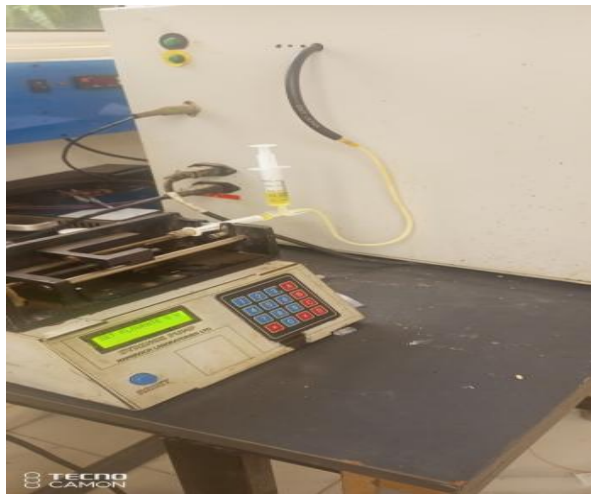

Plate 2: Syringe pump for precursor delivery to the AACVD machine

### 2.2.1 Deposition Parameters

The investigators used the Aerosol-Assisted Chemical Vapour Deposition (AACVD) system to produce AZO thin films by keeping essential parameters under strict control for achieving top-quality and standardized film formation. The constant precursor delivery rate of 0.1 ml/min guaranteed an equal precursor supply that prevented substrate saturation while promoting even decomposition of precursors into films. The study results consistent with Wang et al.'s (2012) findings about improved surface morphology through similar flow rate regulation.

The atomizing voltage reached 6 kV to enable effective aerosol making whereby the precursor solution broke into fine droplets that facilitated uniform surface deposition. The research by Shinde et al. (2011) proves that voltages between 4 kV and 6 kV optimize the atomization process which results in reliable film coverage. Reaction kinetics received optimization at 400°C substrate temperature where material stability together optimized the parameters. The study by Janotti and Van de Walle (2009) indicates that using this temperature leads to enhanced ZnO-based material crystallinity along with reduced defect density.

An optimal deposition process required setting the nozzle-to-substrate distance at 12 mm to avoid shadowing and overspray while ensuring uniform aerosol delivery and film distribution. The authors of Parkin and Palgrave (2005) explain how appropriate distances during AACVD processes function to control film thickness and create superior microstructures. Each sample needed 0.3 ml of precursor solution for controlled thickness production and achievement of targeted AZO layer properties. After conducting preliminary trials scientists chose 0.3 ml of solution because it enabled balanced material growth without causing exceeding deposition.

### 2.2.2 Precursor Solutions

The precursor solutions for Aluminium-doped Zinc Oxide (AZO) thin films were prepared by dissolving Zinc Acetylacetonate (0.15M) and Aluminium Chloride (0.15M) in Methanol (Potter et al, 2018). The solutions were thoroughly mixed to ensure uniformity, resulting in a stable precursor suitable for depositionin conformity with Wang et al., (2012). To maintain consistency and reliability in the deposition process, the same preparation method was applied across all experiments.

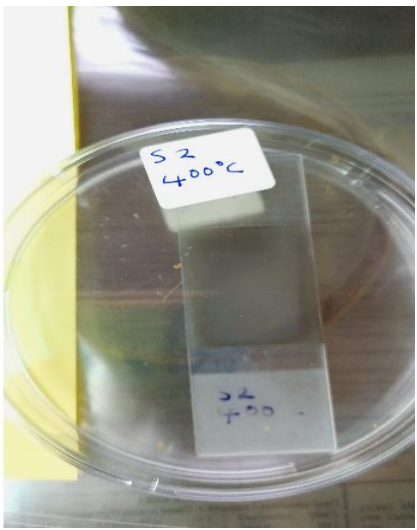

Plate 3: Sample of deposited AZO thin films

### 2.2.3 Substrate cleaning procedure

The soda-lime glass substrates were mildly etched by immersing them in 1M HCl for 24 hours, followed by rinsing with distilled water and drying using a modified centrifuge (Labscience model 800) at 3000 RPM in line with Kołodziejczak-Radzimska & Jesionowski, (2014).

### 2.2.4 Experimental Procedures

In the AACVD system, Zinc Oxide (ZnO) thin films were doped with varying concentrations of Aluminium (Al) by preparing precursor solutions with specific Zn and Al ratios. The Aluminium concentrations (0%, 5%, 10%, 15%, and 20%) were adjusted by modifying the proportion of Aluminium precursor relative to Zinc precursor in the solution. This solution was aerosolized and transported into the heated reaction chamber using nitrogen ($N_2$) as the carrier gas, where it underwent a chemical reaction on the substrate, leading to the deposition of Aluminium-doped ZnO (AZO) thin films. Each Al concentration produced films with distinct properties, enabling a systematic investigation of how varying Al content influences the characteristics of AZO films.

### 2.3.1 Film Deposition

The films were deposited onto soda-lime glass substrates using nitrogen ($N_2$) as the carrier gas. During deposition, the precursor solution was delivered at a controlled flow rate of 0.1 ml/min, while the substrate temperature was maintained at 400°C. Atomization of the precursor was achieved using a voltage of 6 kV, with a nozzle-to-substrate distance of 12 mm. A gas flow rate of 3 L/min was applied to ensure consistent delivery of the precursor aerosol under nitrogen conditions.

To optimize deposition parameters, several critical factors were carefully controlled. The substrate temperature was consistently set at 400°C to ensure uniform film growth, while the gas flow rate of 3 L/min was selected based on preliminary tests to achieve optimal deposition uniformity and film quality. Each sample received 0.3 ml of the precursor solution, and after deposition, the films were annealed at 450°C for 60 minutes in nitrogen. These optimized parameters were chosen to enhance the structural and optical properties of the deposited AZO films.

### 2.3.2 Post-Deposition Optimization

#### 2.3.2.1 Annealing

After deposition, the films were annealed at 450°C for 60 minutes in a nitrogen atmosphere to enhance crystallinity and minimize defects. Annealing at this temperature facilitates atomic reorganization, leading to increased grain size and improved film stability. Minami (2005) highlights that annealing within this range promotes grain growth and enhances the stability of the ZnO lattice structure in doped films. The samples were placed in a tube furnace or annealing chamber designed to withstand high temperatures. To prevent thermal shock, the furnace was gradually heated to 450°C at a controlled ramp-up rate of approximately 5°C to 10°C per minute.

#### 2.3.2.2 Soaking Phase

Once the target temperature was reached, the samples were held at 450°C for 60 minutes. During this soaking phase, key processes such as crystallinity enhancement and defect reduction occurred. The heat facilitated atomic rearrangement, leading to improved grain size and a more stable film structure. Since the annealing was conducted in nitrogen, the process primarily influenced the structural and morphological properties of the AZO films.

#### 2.3.2.3 Cooling Phase

After the annealing process, the samples were gradually cooled to room temperature at a controlled rate of approximately 5°C to 10°C per minute. This slow cooling minimized thermal stress, helping to preserve the improved structural and electrical properties of the films while preventing the formation of new defects.

#### 2.3.2.4 Annealing Gas

Nitrogen ($N_2$) was used as the annealing gas for the Aluminium-doped Zinc Oxide (AZO) films in the nitrogen series (S1N to S5N). Annealing in nitrogen provided an inert environment, preventing oxidation and allowing for a comparative analysis of how a non-oxidizing atmosphere influences the structural and functional properties of AZO films. This comparison is crucial for understanding the role of annealing conditions in optimizing the material's characteristics for specific applications.

### 2.3.3 Thickness Measurement

A stylus profilometer was used for thickness measurement, providing high-precision scanning across multiple points to determine an average film thickness with an error margin of ±2 nm. Bhatia et al. (2020) highlight the reliability of stylus profilometry for thin-film measurements, particularly when accuracy is essential for optical applications.

## 2.4 Advanced Characterization of AZO Thin Films

Characterization techniques included X-ray Diffraction (XRD) and Ultraviolet-Visible (UV-Vis) Spectroscopy. XRD was employed to analyze crystallinity and phase structure, confirming the incorporation of Aluminium and ensuring phase purity in AZO films. Ozgur et al. (2005) emphasize XRD's effectiveness in detecting variations in crystal lattice parameters, which is crucial for assessing the impact of doping and nitrogen as a carrier gas.

UV-Vis spectroscopy was used to evaluate optical properties, including transmittance and band gap, offering insights into the film's transparency and potential for optoelectronic applications. This approach aligns with the findings of Kołodziejczak-Radzimska and Jesionowski (2014), which highlight the relationship between bandgap energy and doping levels. Together, these characterization techniques provide a comprehensive framework for assessing and optimizing AZO thin films for advanced applications.

### 2.5 Data Analysis and Interpretation

Data analysis involved both structural and optical assessments of the Aluminium-doped Zinc Oxide (AZO) films. X-ray Diffraction (XRD) data were used to analyze crystallite size and peak intensity, offering insights into the films' crystallinity and phase structure. Optical properties were evaluated through UV-Visible spectroscopy (UV-Vis), enabling comparisons of transmittance and band gap energies. Furthermore, crystallite size and dislocation density were calculated to assess the structural integrity and quality of the films, providing a thorough understanding of their performance.

## 3. RESULTS AND DISCUSSION

### 3.1 Results

#### 3.1.1 Structural Properties of AZO Films

##### 3.1.1.1 XRD pattern for undoped ZnO

Figures 3.1 show the XRD patterns for undoped ZnO (AZO) with nitrogen as the carrier gas.

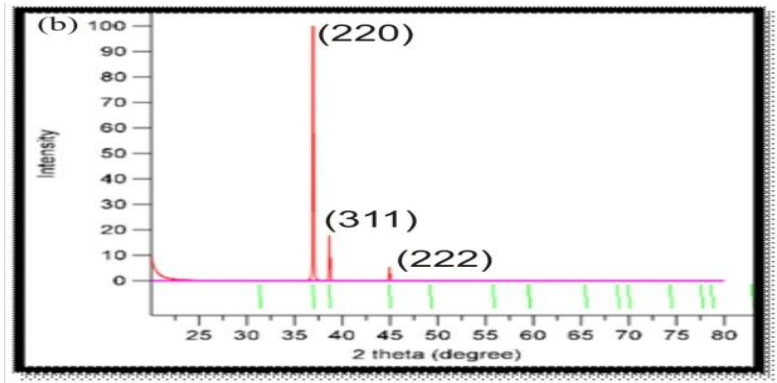

*Fig 3.1: XRD pattern for undoped ZnO (Nitrogen as carrier gas)*

##### 3.1.1.2 XRD pattern for Al-doped ZnO

The XRD patterns presented in Figure 3.2 to figure 3.5 illustrate the structural characteristics of Aluminium-doped Zinc Oxide (Al-doped ZnO) films at varying doping levels (5%, 10%, 15%, and 20%) and with carrier gases nitrogen.

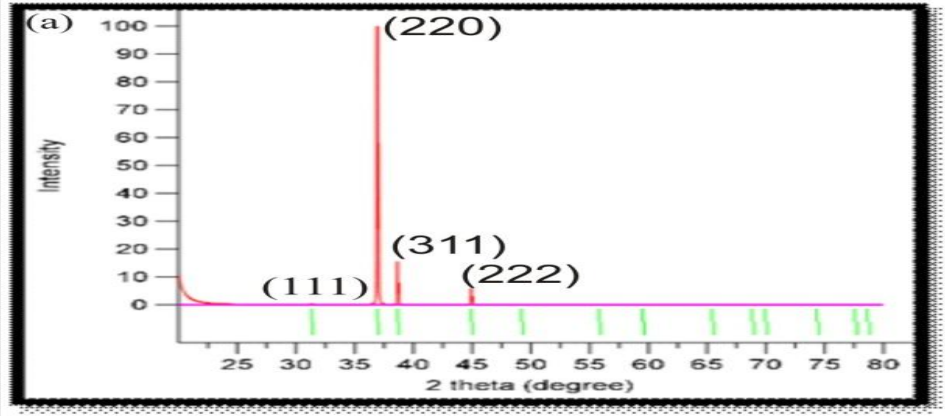
Fig 3.2: XRD pattern for 5% Al-doped ZnO (Nitrogen as carrier gas)

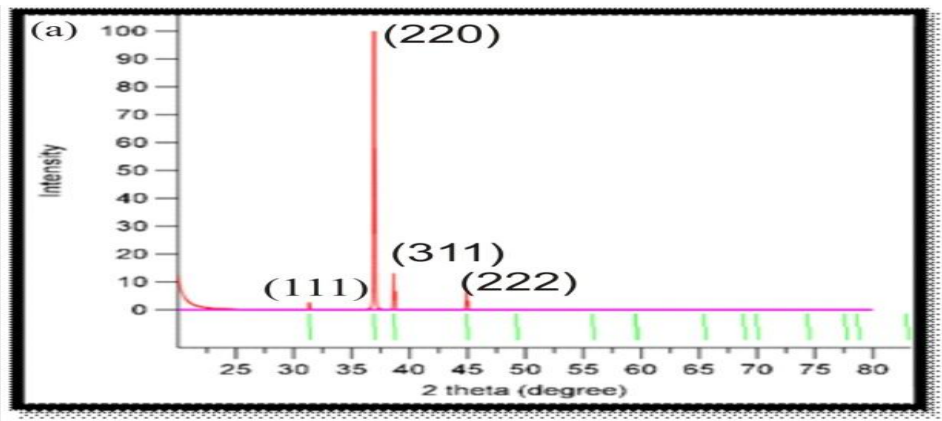
Fig 3.3: XRD pattern for 10% Al-doped ZnO (Nitrogen as carrier gas)

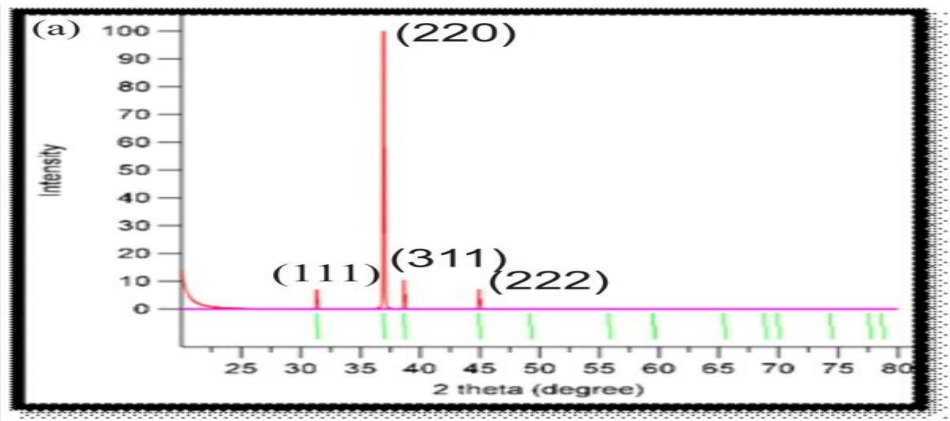
Fig 3.4: XRD pattern for 15% Al-doped ZnO (Nitrogen as carrier gas)

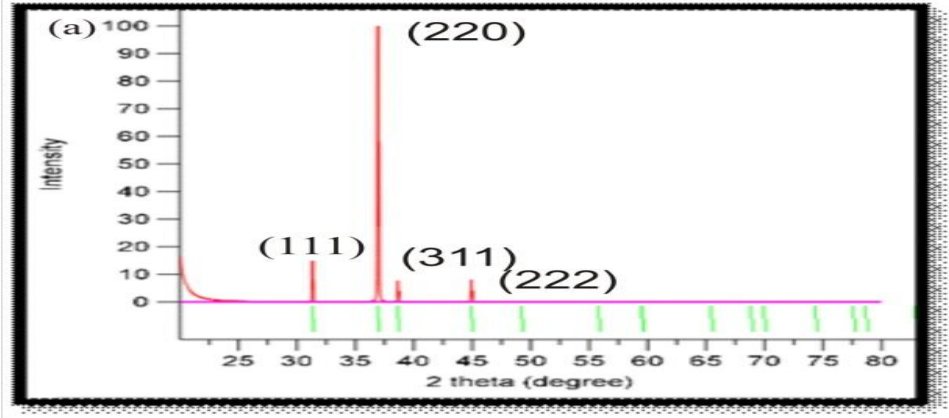

Fig 3.5: XRD pattern for 20% Al-doped ZnO (Nitrogen as carrier gas)

### 3.1.2 Crystallite Size and Dislocation Density of Aluminium-doped Zinc Oxide (AZO) Thin Films

Tables 3.1 shows the calculated crystallite sizes and dislocation densities of Aluminium-doped ZnO (AZO) samples at varying doping levels (0%, 5%, 10%, 15%, and 20%) using nitrogen as carrier gas.

**Crystallite Size**

Using Scherrer equation: $D = (0.9 \times \lambda)/(\beta \times \cos\theta)$  (3.1)

**Dislocation Density**

$\delta = 1/D^2$  (3.2)

**FWHM in Degrees (XRD):**

$FWHM = \dfrac{K.\lambda}{D.\cos(\theta)}$  (3.3)

**Table 3.1: Crystallite Size and Dislocation Density (Nitrogen as carrier gas)**

| Al Content (%) | Peak (hkl) | 2θ (degrees) | FWHM (degrees) | Crystallite Size (nm) | Dislocation Density ($10^{15}$ lines/m²) |
|---|---|---|---|---|---|
| 0 | (220) | 31.7 | 0.24 | 34.8 | 0.825 |
|   | (311) | 34.4 | 0.25 | 33.6 | 0.885 |
|   | (222) | 36.2 | 0.25 | 33.7 | 0.88 |
| 5 | (111) | 31.7 | 0.26 | 32.1 | 0.97 |
|   | (220) | 34.4 | 0.27 | 31.1 | 1.03 |

|    | (311) | 36.2 | 0.27 | 31.2 | 1.03 |
|    | (222) | 47.5 | 0.3  | 28.7 | 1.21 |
| 10 | (111) | 31.7 | 0.28 | 29.8 | 1.12 |
|    | (220) | 34.4 | 0.29 | 28.9 | 1.2  |
|    | (311) | 36.2 | 0.29 | 29   | 1.19 |
|    | (222) | 47.5 | 0.32 | 26.9 | 1.38 |
| 15 | (111) | 31.7 | 0.3  | 27.8 | 1.29 |
|    | (220) | 34.4 | 0.31 | 27   | 1.37 |
|    | (311) | 36.2 | 0.31 | 27.2 | 1.35 |
|    | (222) | 47.5 | 0.34 | 25.3 | 1.56 |
| 20 | (111) | 31.7 | 0.32 | 26.1 | 1.47 |
|    | (220) | 34.4 | 0.33 | 25.4 | 1.55 |
|    | (311) | 36.2 | 0.33 | 25.5 | 1.54 |
|    | (222) | 47.5 | 0.36 | 23.9 | 1.75 |

Figures 3.6 and 3.7 illustrate the relationship between Aluminium doping concentration and the microstructural properties; crystallite size and dislocation density of AZO thin films with nitrogen as carrier gas.

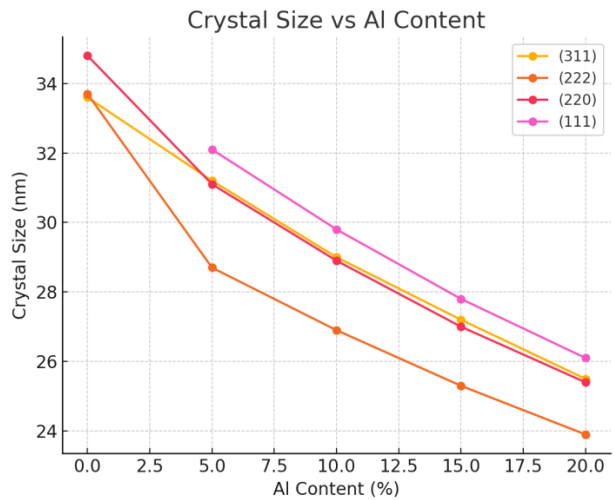

*Fig 3.6: Crystallite Size versus Al Doping Concentration for AZO Thin Films (Nitrogen as carrier gas)*

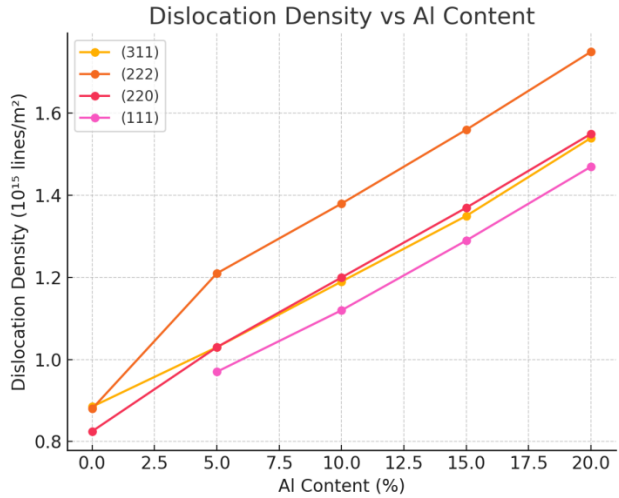

Figure 3.7: Dislocation Density versus Al Doping Concentration for AZO Thin Films (Nitrogen as carrier gas)

### 3.1.3 Profilometry analysis of Aluminium-doped Zinc Oxide (AZO) Thin Films

Table 3.2 showing the measured thickness of the AZO thin films using profilometry was found to increase slightly with increasing doping concentration for nitrogen carrier gas.

**Table 3.2: Film Thickness of AZO Thin Films at Various Doping Levels and Carrier Gases**

| Doping Concentration | Nitrogen Carrier Gas (nm) |
|---|---|
| 0% Al (ZnO) | 102 |
| 5% Al | 107 |
| 10% Al | 110 |
| 15% Al | 112 |
| 20% Al | 115 |

Figure 3.8, shows that with higher doping levels, variations in deposition conditions, such as temperature fluctuations or gas flow rates, may introduce minor inconsistencies. Measurement errors from the profilometer were estimated to be within ±5 nm, but these do not significantly affect the observed trend.

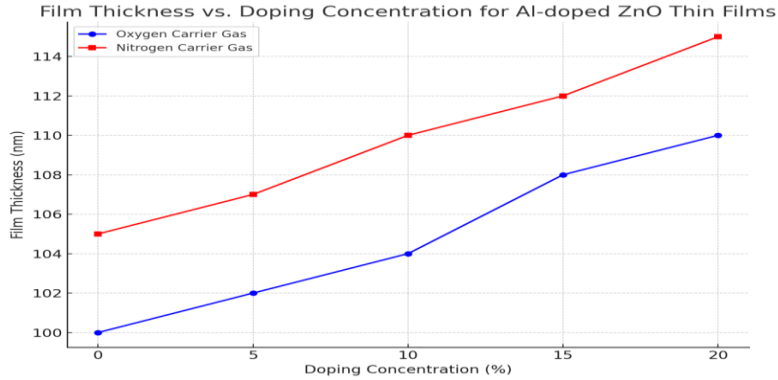

Fig 3.8: Film Thickness versus Al-Doping Concentration of ZnO Thin Films (Nitrogen/Oxygen as carrier gas)

### 3.1.3 Optical Properties
### 3.1.3.1 UV-Vis spectra of AZO films.

Figure 3.9 shows the UV absorption spectra of undoped ZnO using nitrogen ($N_2$) and oxygen ($O_2$) as carrier gases show nearly identical behaviour across the measured wavelength range.

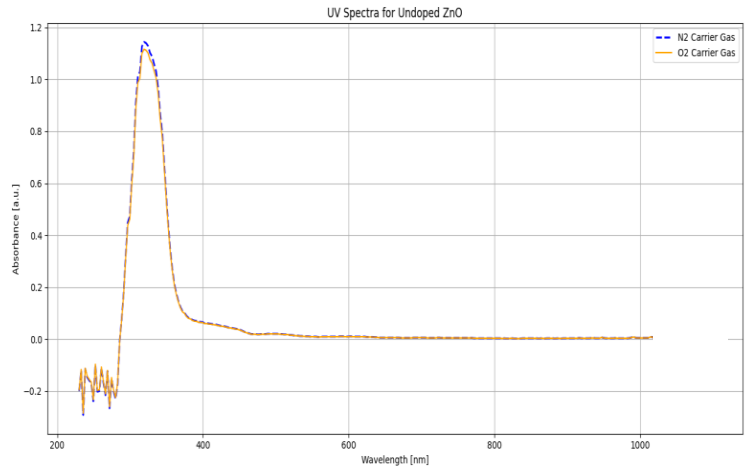

Fig 3.9 UV-Vis Absorbance versus Wavelength for Undoped ZnO at 0% Concentration (Nitrogen as Carrier Gas)

Figure 3.10 below shows the UV absorption spectra of Al-doped ZnO with Nitrogen, as the carrier gas exhibit significant changes with varying doping levels (5%, 10%, 15%, and 20%). All samples show a sharp rise in absorbance around 365nm, corresponding to the band-gap absorption of ZnO.

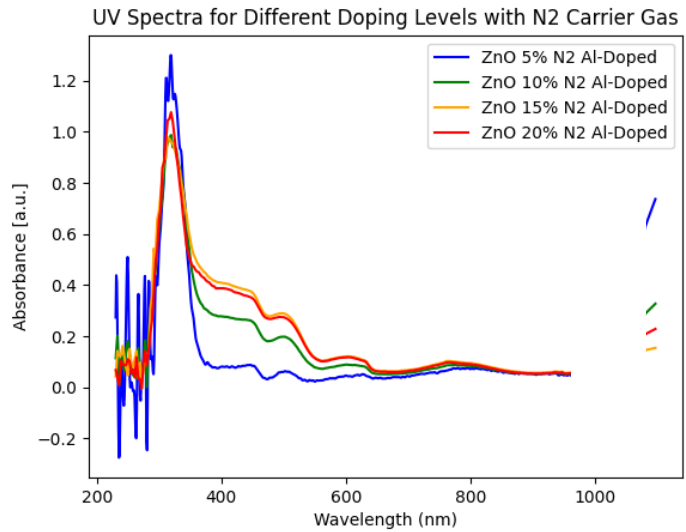

Fig. 3.10: UV-Vis Absorbance versus Wavelength for Al-Doped ZnO at Various Concentrations (Nitrogen as Carrier Gas)

### 3.1.3.2 Bandgap Analysis

From the calculated values from equation 2.16, the various absorbance coefficients $(\alpha h\nu)^2$ were plotted against the photon energy $h\nu$ as shown in figures 3.11 to 3.15 for the Al-doped Zinc Oxide (AZO) thin film samples at different doping levels (0%, 5%, 10%, 15%, and 20%) and the tangent of the graphs at the x-axis give uses the energy bandgap for AZO with Oxygen and Nitrogen as carrier gases respectively.

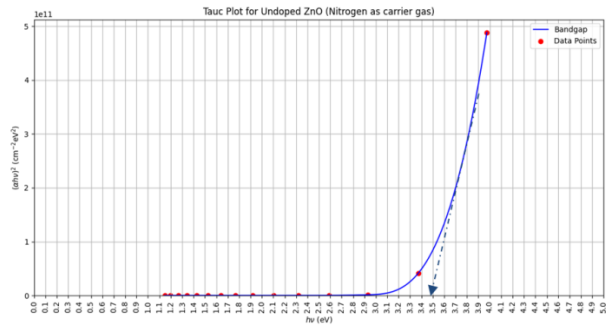

Fig 3.11: Tauc plot of $(\alpha h\nu)^2$ versus $h\nu$ for Undoped ZnO (Nitrogen as carrier gas)

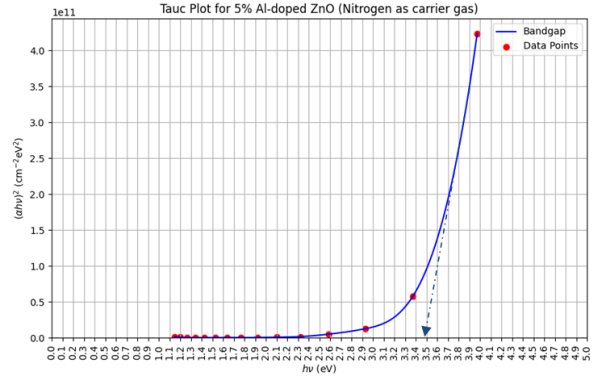

Fig 3.12: Tauc plot of $(\alpha h\nu)^2$ versus $h\nu$ for ZnO doped with Al at 5% doping level (Nitrogen as carrier gas)

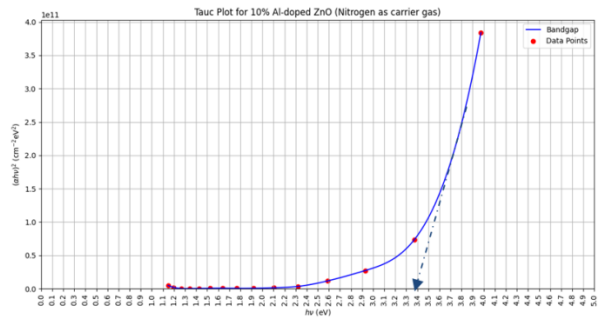

Fig 3.13: Tauc plot of $(\alpha h\nu)^2$ versus $h\nu$ for ZnO doped with Al at 5% doping level (Nitrogen as carrier gas)

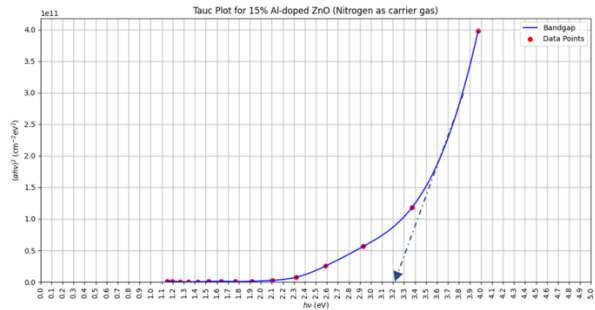

Fig 3.14: Tauc plot of $(\alpha h\nu)^2$ versus $h\nu$ for ZnO doped with Al at 5% doping level (Nitrogen as carrier gas)

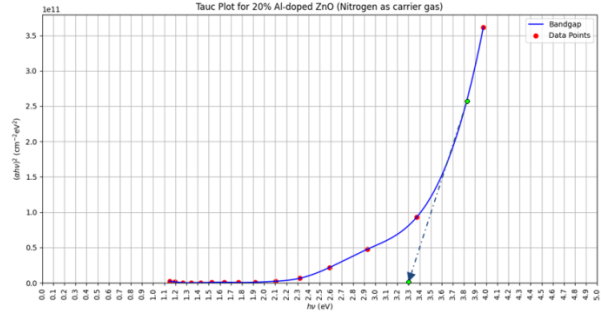

Fig 3.15: Tauc plot of (αhν)² versus hν for ZnO doped with Al at 5% doping level (Nitrogen as carrier gas)

## 3.2 DISCUSSIONS

### 3.2.1 The influence of carrier gas on crystallinity and phase composition.

Figure 3.1 presents the XRD patterns for undoped ZnO using nitrogen as the carrier gas. A prominent peak appears around 37-38° (2θ) for the (220) plane, indicating good crystallinity. The polycrystalline nature of ZnO is further confirmed by smaller peaks at 32° and 43°, corresponding to the (311) and (222) planes. The choice of nitrogen ($N_2$) as a carrier gas significantly influences the crystallinity and phase composition of Al-doped ZnO (AZO) films.

Figures 3.2 to 3.5 illustrate the XRD patterns for Al-doped ZnO at different doping levels under nitrogen. A dominant peak around 37-38° (2θ) for the (220) plane is consistently observed, along with additional peaks for the (111), (311), and (222) planes. Nitrogen as a carrier gas tends to produce sharper, more intense peaks, suggesting enhanced crystallinity in the deposited films.

### 3.2.2 Crystallite Size and Dislocation Density

Figures 3.6 and 3.7, along with Table 3.1, summarize the trends in crystallite size and dislocation density for Al-doped ZnO deposited using nitrogen as the carrier gas. As the Al content increases from 0% to 20%, the crystallite size gradually decreases, while the dislocation density rises, indicating increased lattice imperfections. The use of nitrogen as a carrier gas results in smaller crystallite sizes and higher dislocation densities, influencing the structural properties of the AZO films.

### 3.2.2 Impact of carrier gas on optical absorbance.

In general, the use of nitrogen as a carrier gas does not significantly affect the optical transmittance of undoped ZnO. As shown in the UV-Vis absorbance spectra (Fig. 3.9), undoped ZnO exhibits a sharp rise in absorbance near 300 nm, with a peak around 365 nm corresponding to ZnO's band-gap absorption. After this peak, absorbance drops rapidly, indicating high transparency in the visible and infrared regions.

However, when Al doping is introduced, nitrogen as a carrier gas influences the intensity and range of absorption. For instance, at a 5% doping level, AZO films deposited with nitrogen show a peak absorbance of 1.3 units (Fig. 3.10). This suggests that nitrogen leads to slightly higher UV absorbance. The effect becomes more pronounced at higher doping levels (15% and 20%), where AZO films exhibit a broader absorption band in the visible and infrared regions, indicating increased optical absorption beyond the UV range.

### 3.2.3 Band gap energy

Figure 3.11 presents the bandgap variations of ZnO films as determined from Tauc plots, while Figures 3.12 to 3.15 illustrate the bandgap variations of AZO films with 5%, 10%, 15%, and 20% Aluminium doping concentrations. These values were derived from Tauc plots of the absorbance coefficient $(\alpha h v)^2$ plotted against photon energy ($hv$).

For undoped ZnO, the bandgap energy is measured at 3.21 eV when nitrogen is used as the carrier gas. As Aluminium doping increases, the bandgap also increases. At 5% Al doping, the bandgap rises to 3.26 eV, followed by 3.29 eV at 10% doping. With 15% doping, the bandgap further increases to 3.31 eV, and at 20% Al doping, it reaches 3.33 eV. These trends, shown in Figures 3.12 to 3.15, indicate that Aluminium incorporation leads to a progressive widening of the bandgap, with nitrogen as the carrier gas influencing the final energy values.

### 3.2.4 Thin Film Thickness

The Al-doped ZnO thin films reached a thickness control within ±5 nm range as the doping concentration increased while using nitrogen carrier gas made the films slightly thicker. The kinetic study of Al-doped Zinc Oxide (AZO) thin films at different doping concentrations is presented in Table 3.2 along with Figure 3.8. A gradual increase exists between film thickness and dopant levels based on the obtained data. The film grows from 102 nm at 0% Al doping to 115 nm when doping concentration reaches 20%. The growth patterns of films are modified by nitrogen during their developmental phase.

### 3.3 Unique Findings of This Study

The research demonstrates that nitrogen carrier gas produces slightly thicker films which grew smaller crystallites because the grains experienced constrained development. Structural defects increase because nitrogen introduces denser dislocations to the material. The UV-visible absorption spectra demonstrate that films annealed with nitrogen exhibit better ultraviolet light absorption when the aluminium content reaches 5% concentration. Cyclic changes in bands gap energies follow higher Aluminium content rates until 20% Al-doped ZnO which shows a bandgap rise of 0.1 eV from undoped ZnO to generate the Burstein-Moss effect. The absorption spectrum extends into the visible wavelength range as a result of integrating 15% and 20% Al doping into thefilms. The extended optical activity into both UV and visible ranges makes these films promising candidates for improved optoelectronic applications.

### 3.4 Discussion on Mechanism

AZO film properties depend mostly on the amount of Al doping. The XRD findings show that aluminium doping produces lattice strain through Zn atom substitutions which triggers crystallite size decrease and densified dislocation density. Ganesh et al (2022) supports this finding because increasing Al levels in the material creates more extensive lattice distortions that block further grain growth. The extent of UV light absorption increases slightly in nitrogen-annealed samples because of the higher concentration of absorption centers related to defects according to Lee et al, (2004).

The Burstein-Moss effect functions as the major mechanism responsible for expanding bandgaps. The introduction of carriers from Al doping increases the Fermi level which widens the bandgap of ZnO from its base value of 3.21 eV to 3.37 eV for samples with 20% Al content. The presence of high Al concentration induces absorption expansion in the visible spectrum and infrared range because of the defect states and impurity levels that Al doping creates. Added broad absorption expands the films' wavelength sensitivity mostly in the visible spectrum range according to Singh et al. (2021).

**3.5 Implications for the optimization of AACVD processes.**

Research outcomes show key factors for AACVD process enhancement that need attention concerning the selection of carrier gases and dopant concentrations. The utilization of nitrogen as carrier gas leads to both shorter crystallite sizes and heavier film thickness along with higher densities of dislocations in comparison with oxygen usage. The research indicates that nitrogen limits grain expansion while creating additional structural abnormalities during growth which benefits UV absorption at low doping concentrations. Oxygen presence creates improved crystalline properties together with larger grain dimensions which match well with product requirements that need high structural quality combined with less flaws. Oxygen proves superior for maximizing crystal quality but nitrogen provides better properties regarding defects thus following Singh et al, (2023).

The Burstein-Moss effect enables modifications of the bandgap through Aluminium doping which allows bandgap tuning for optimized optical characteristics. At Al doping levels between 15-20% the absorption spectrum expands which enables the films to become viable for visible light applications in solar cell and photodetector designs. The AACVD process allows operators to control doping levels together with carrier gas composition which enables production of different material properties for industrial purposes as reported by Li et al, (2018).

**Conclusion**

By using Aerosol-Assisted Chemical Vapour Deposition (AACVD) with nitrogen ($N_2$) as the carrier gas laboratory personnel succeeded in synthesizing Aluminium-doped Zinc Oxide (AZO) thin films while conducting systematic investigations of aluminium (Al) doping concentration variations between 0%, 5%, 10%, 15%, and 20% to analyze structural optical and morphological film properties. The use of nitrogen carrier gas during AZO film synthesis proves essential for modifying both their crystallinity level as well as defect density and optical characteristics thus making it an essential parameter for developing materials for specific optoelectronic device applications.

X-ray Diffraction (XRD) analysis showed that nitrogen promotes better crystallinity through the appearance of distinctive strong peaks representative of the (220), (311), and (222) planes. An increase in Al doping levels resulted in smaller crystal size while increasing the number of dislocations which generated higher structural defects. The introduction of Al doping in crystal structures produces defects that interfere with grain growth while nitrogen maintains crystallinity. Profilometry measurements showed that films became thicker as Al doping increased from 102 to 115 nm starting from undoped ZnO up to 20% Al-doped ZnO thereby confirming the influence of nitrogen on film growth dynamics.

UV-Vis spectroscopy showed nitrogen-annealed films reach better UV absorbance values at 5% doping concentration and the bandgap energy increases from 3.21 eV in undoped ZnO to 3.33 eV in 20% Al-doped ZnO. The Burstein-Moss effect clarifies this bandgap enlargement because high free carrier numbers raise the Fermi level into the conduction band. The visible and infrared region absorption expanded in the Al-doped films when the doping concentration reached 15% and 20%. The defect states and impurity levels resulting from Al doping cause the films to expand their sensitivity range for both UV and visible wavelengths.

This research demonstrates that selecting the right carrier gas and optimizing doping concentration is vital to achieve optimized properties in AZO thin films which function in optoelectronic systems. Thin films produced with nitrogen carrier gas yield smaller crystallites and denser dislocations and additional thickness, thus meeting requirements for better UV absorption capabilities with defect-sensitive characteristics. Aplications requiring high crystallinity and structural quality usually benefit from the use of oxygen as the carrier gas. The AACVD method demonstrates versatile properties since it enables precise adjustments of doping levels and carrier gases which controls AZO film properties.